\documentclass[sigconf, nonacm]{acmart}

\newcommand{\eat}[1]{}

\usepackage{latexsym}
\usepackage{amsfonts}
\usepackage{amsmath}
\usepackage{xcolor}
\usepackage{colortbl}
\usepackage{epsfig}
\usepackage{xspace}
\usepackage{graphicx}
\usepackage{subfigure}
\usepackage{paralist}
\usepackage{enumerate}
\usepackage[color,matrix,arrow,all]{xy}
\usepackage{comment}
\usepackage{booktabs}
\usepackage{balance}
\usepackage{stmaryrd}
\usepackage{pifont}
\usepackage{hhline}
\usepackage{listings}
\usepackage{array}
\usepackage{float}
\usepackage[flushleft]{threeparttable}

\usepackage{mathrsfs}
\usepackage{makecell}
\usepackage{xparse}
\usepackage{wrapfig}

\usepackage{epsfig}
\usepackage{multirow}
\usepackage{url}

\usepackage{multirow}
\usepackage{natbib}
\usepackage{graphicx}

\usepackage{listings}
\usepackage{framed}
\usepackage{xcolor}
\usepackage{color}
\usepackage{geometry}
\usepackage{changepage}
\colorlet{shadecolor}{gray!20}

\definecolor{shadecolor}{RGB}{220,220,220}

\definecolor{inputcolor}{RGB}{255,139,35}
\definecolor{outputcolor}{RGB}{120,212,252}
\definecolor{embedcolor}{RGB}{254,127,156}
\definecolor{maskcolor}{RGB}{122,128,255}
\definecolor{ecolor}{RGB}{58,149,54}

\definecolor{highcolor}{RGB}{255,153,153}
\definecolor{midcolor}{RGB}{255,204,204}
\definecolor{lowcolor}{RGB}{204,229,255}

\newtheorem{example}{Example}

\usepackage{tikz}
\usetikzlibrary{shapes,snakes}
\usetikzlibrary{calc}

\usepackage{algorithm} 
\usepackage{algorithmic}  
\usepackage[algo2e]{algorithm2e} 

\usepackage[export]{adjustbox}

\definecolor{green}{RGB}{0,128,0}

\definecolor{yellow}{RGB}{255,200,18}

\newcommand{\add}[1]{\textcolor{blue}{{#1}}}

\newcommand{\at}[1]{\protect\ensuremath{\mathsf{#1}}}
\newcommand{\att}[1]{\textbf{\texttt{#1}}}

\newcommand{\stab}{\vspace{1.2ex}\noindent}
\newcommand{\sstab}{\rule{0pt}{8pt}\\[-2.2ex]}

\newcommand{\ra}{\rightarrow}

\newcommand{\bi}{\begin{itemize}}
\newcommand{\ei}{\end{itemize}}

\newcommand{\be}{\begin{enumerate}}
\newcommand{\ee}{\end{enumerate}}
\newcommand{\beqn}{\begin{eqnarray*}}
\newcommand{\eeqn}{\end{eqnarray*}}

\newcommand{\stitle}[1]{\stab\noindent{\bf #1}}
\newcommand{\etitle}[1]{\vspace{1mm}\noindent{\underline{\em #1}}}

\newcommand{\eg}{{\em e.g.,}\xspace}

\newcommand{\eop}{\hspace*{\fill}\mbox{$\Box$}}     

\newcommand{\sys}{\att{VerifAI}\xspace}
\newcommand{\chat}{\texttt{ChatGPT}\xspace}

\makeatletter
    \newcommand\figcaption{\def\@captype{figure}\caption}
    \newcommand\tabcaption{\def\@captype{table}\caption}
\makeatother

\tikzstyle{mybox} = [draw=black, fill=black!5, thick,
   rectangle, rounded corners, inner sep=0pt, inner ysep=6pt]
\tikzstyle{fancytitle} =[fill=black, text=white]

\NewDocumentCommand{\nan}{ mO{} }{\textcolor{blue}{\textsuperscript{\textit{Nan}}\textsf{\textbf{\small[#1]}}}}

\NewDocumentCommand{\yang}{ mO{} }{\textcolor{green}{\textsuperscript{\textit{yang}}\textsf{\textbf{\small[#1]}}}}
\usepackage{balance}

\begin{document}

\title{VerifAI: Verified Generative AI}

\author{Nan Tang}
\affiliation{
  \institution{HKUST (GZ)} 
  \country{China}
}

\author{Chenyu Yang}
\affiliation{%
  \institution{HKUST (GZ)} 
  \country{China}
}

\author{Ju Fan}
\affiliation{%
  \institution{Renmin Univ.} 
  \country{China}
}

\author{Lei Cao}
\affiliation{%
  \institution{CSAIL, MIT} 
  \country{USA}
}

\author{Yuyu Luo}
\affiliation{%
  \institution{HKUST (GZ)} 
  \country{China}
}

\author{Alon Halevy}
\affiliation{%
  \institution{Meta AI} 
  \country{USA}
}

\begin{abstract}
Generative AI has made significant strides, yet concerns about the accuracy and reliability of its outputs continue to grow. Such inaccuracies can have serious consequences such as inaccurate decision-making, the spread of false information, privacy violations, legal liabilities, and more. Although efforts to address these risks are underway, including explainable AI and AI regulation practices such as transparency, privacy protection, bias mitigation, and social and environmental responsibility, misinformation caused by generative AI will remain a significant challenge. We propose that verifying the outputs of generative AI from a data management perspective is an emerging issue for generative AI. This involves analyzing the underlying data from multi-modal data lakes, including text files, tables, and knowledge graphs, and assessing its quality and consistency. By doing so, we can establish a stronger foundation for evaluating the outputs of generative AI models. Such an approach can ensure the correctness of generative AI, promote transparency, and enable decision-making with greater confidence. Our vision is to promote the development of verifiable generative AI and contribute to a more trustworthy and responsible use of AI.
\end{abstract}

\maketitle

\section{Introduction}
\label{sec:intro}

Making effective decisions based on data requires data of high quality. The exact meaning of high quality varies from one application to another, but typically it means that the data has undergone verification, validation, and evaluation to ensure its reliability and accuracy for a specific use case or application. Furthermore, we may also require that the appropriate levels of security and privacy protection be in place. This is particularly crucial in fields such as finance, healthcare, and government, where the decisions based on the data can affect individuals and society as a whole. Organizations that rely on good data for making important decisions must take steps to ensure that the data they use is trustworthy and reliable.

Regrettably, real-world data is often incomplete, inconsistent, or inaccurate. Researchers across different fields have invested significant effort in addressing these issues. For instance, the database community has developed methods for detecting data errors~\cite{DBLP:journals/pvldb/AbedjanCDFIOPST16}, while the natural language community has developed techniques for identifying fake news and misinformation~\cite{DBLP:journals/corr/abs-2205-04274}.

The issue of data quality becomes more important in the era of generative AI, where advanced generative models like \chat, Midjourney, and SlidesAI can generate intricate outputs. This technology has immense potential to transform various fields, including relational data synthesis, natural language question answering, image generation for advertising and marketing, and many others.

More specifically, 
numerous companies have audacious and transformative plans for integrating generative AI into a wide range of their products, including Microsoft (Excel, Word, PowerPoint, Bing Search), Google, Meta, Alibaba, and Baidu, and it is reasonable to expect  this trend to proliferate beyond the big technology companies.   Consequently, the data produced by these systems will be utilized for multiple purposes, such as decision-making, knowledge acquisition, and presentations. However, it is crucial to acknowledge that the accuracy and reliability of the generated data cannot be guaranteed. Therefore, it is vital to exercise caution when utilizing such data to make important decisions or draw conclusions.

There are significant risks associated with using ``bad'' data generated by generative AI, such as the following.

\be
    \item {\em Inaccurate decision-making:} 
    Decisions based on bad data can lead to poor outcomes, \eg financial losses. 

    \item {\em Spreading misinformation:} 
    Bad data can be utilized to disseminate false information, and individuals may rely on the knowledge from generative AI irrespective of its accuracy.

    \item {\em Privacy violations:} 
    It can lead to privacy violations if personal information is shared or used inappropriately.

    \item {\em Legal liabilities:} 
    Organizations that make decisions based on bad data can face legal liabilities if they cause harm to individuals or violate laws and regulations.

    \item {\em Damage to trust:} The use of bad data can erode trust in organizations, leading to long-term reputational damage.
\ee

Although the risks discussed above are already prevalent when using low quality data before the era of generative AI, they are  significantly amplified with the widespread adoption of generative models. In addition to the many discussions about regulating generative AI, from a technical perspective there have been two main bodies of work on the topic, explainable AI and AI regulations.

\begin{figure*}[t!]
\begin{center}
\vspace{-1em}
\includegraphics[width=0.9\textwidth]{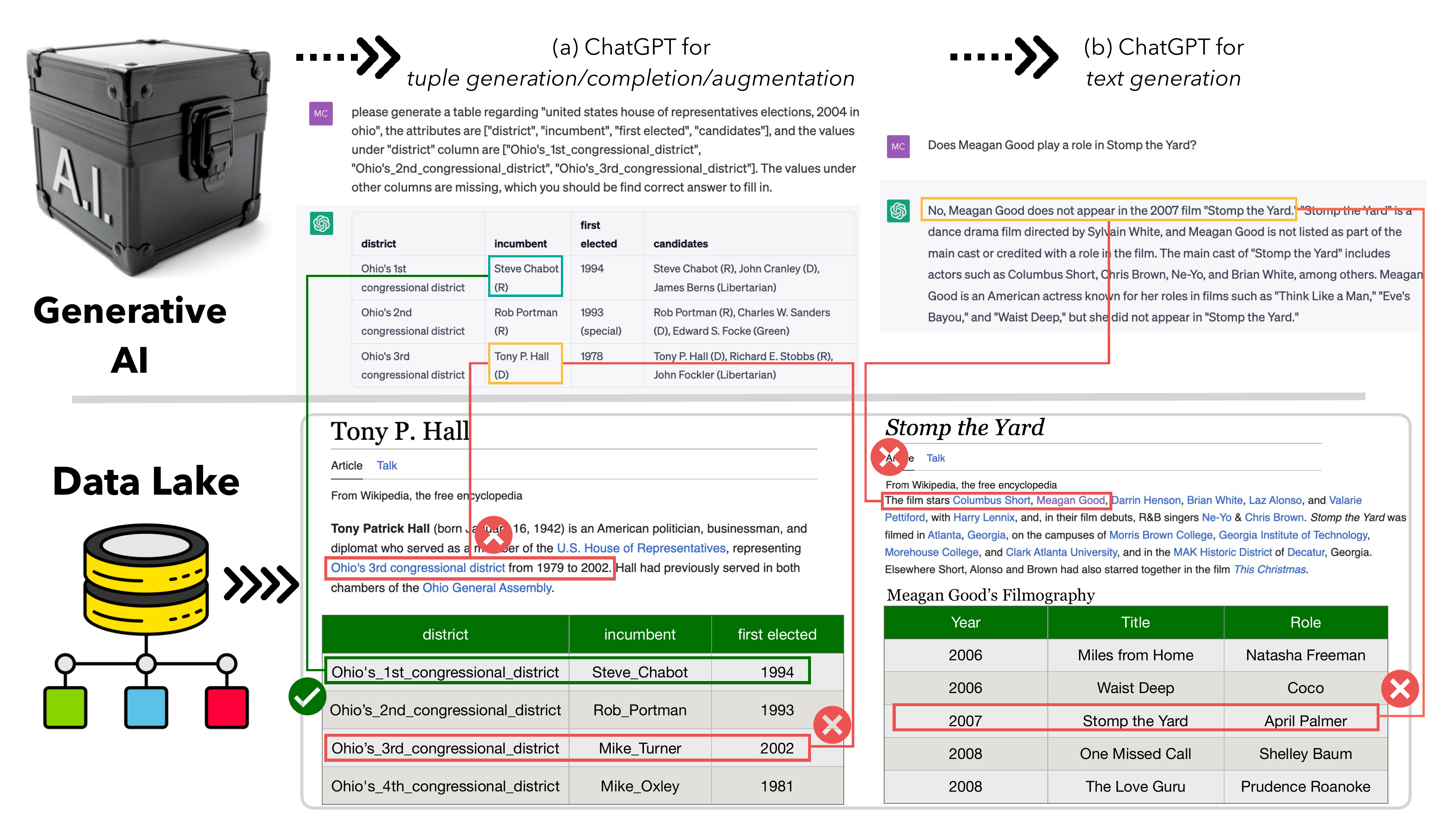}
\end{center}
\vspace{-1em}
\caption{Generative AI can generate (a) values in tuples and (b) text. Our system, \sys, tries to either verify or refute generated value, by reasoning the (generated data, evidence) pair where the evidence is discovered from data lakes.}
\label{fig:examples}
\vspace{-1em}
\end{figure*} 

Explainable AI (XAI)~\cite{e23010018} seeks to provide explanations for the outputs produced by AI systems. While XAI can be a valuable tool for improving the trustworthiness of generative AI, it cannot completely eliminate concerns regarding the accuracy and reliability of generated outputs. This is due in part to the fact that generative AI models can be highly complex, which makes it difficult to provide a complete explanation of how a given output was produced.


AI regulations involve the ethical and accountable development and use of AI systems, which encompasses a range of practices such as transparency, privacy protection, bias mitigation, and social and environmental responsibility. Although model providers like OpenAI and Google are making efforts to regulate the development of generative AI models, the issue of flawed generative data is expected to continue to be a significant challenge.

\stitle{Our Vision.}
We advocate for the establishment of a more reliable framework to evaluate the accuracy and reliability of generative AI outputs. We propose a data management perspective for verifying generative AI outputs that is based on identifying supporting data from data lakes and reasoning with it in order to assess the quality and consistency of the generated data.
We believe that a verification approach can complement the aforementioned approaches in order to ensure that systems using generative AI  are deployed responsibly, effectively, and in a trustworthy manner. 
It is important to clarify that our proposed approach focuses on verifying generative data that has ground truth, as opposed to subjective data~\cite{DBLP:conf/enase/Halevy20}. 

We illustrate our approach with the following examples based on running our proposed framework \sys.

\begin{example}
\label{exam:gen}
$[$Tuple generation/completion/augmentation (Figure~\ref{fig:examples}(a)).$]$
Consider the prompt that provides a serialized table with missing values in attribute \att{incumbent} and asks \chat to complete these tuples. 
The new table with completed tuples by \chat is shown below the prompt.

For the first tuple, \sys can ``search'' a tuple in the data lake that verifies the imputed and generative value that the \att{incumbent} value is correct.

For the third tuple, \sys can search a tuple and a text file in the data lake that both validate the imputed value to be incorrect.

\sstab
$[$Text generation (Figure~\ref{fig:examples}(b)).$]$
We asked \chat a question ``Does Meagan Good play a role in Stomp the Yard'' and got an answer as shown in the figure.

\sys can search a text file and a tuple in the data lake that both validate the generated text to be incorrect.
\eop
\end{example}

The examples above only scratch the surface of generated (bad) data. These data can be leveraged for downstream applications such as analytical (\eg OLAP) queries and data visualization, as well as query acquisition. However, if not properly managed, generative data can have negative and even disastrous consequences.

Despite efforts from model providers (\eg OpenAI) and retrieval-enhanced methods~\cite{DBLP:journals/corr/abs-2112-04426}
for improving the accuracy of generative AI during the data generation process, the spread of misinformation from generative AI will remain a significant problem.
Our proposed post-generation verification approach can also complement these methods by enhancing the overall reliability of generative AI. By combining our verification approach with those existing and rapidly evolving generative methods, we can create a synergistic effect that further improves the accuracy and reliability of generative AI.

\stitle{Challenges.}
Generative AI can generate data in various contexts and domains, utilizing the {\em world knowledge} that the model has learned~\cite{DBLP:journals/corr/abs-2112-04426}. However, when the generated data is dirty or inaccurate, traditional data cleaning and integration methods that rely on the data at hand may not be sufficient. Moreover, when generative AI is used in the context of a specific enterprise, we may need to consult data that is specific to the enterprise in order to verify the correctness of the generated data, raising several new challenges.

{\bf (C1)} \textit{Indexing and searching multi-modal data lakes.}
Although there have been efforts to manage data lakes with relational data~\cite{DBLP:conf/icde/FernandezAKYMS18} and textual data~\cite{colbert}, indexing multi-modal data lakes at scale and effectively retrieving top-$k$ data instances cross data modalities remains an unsolved problem.

{\bf (C2)} \textit{Cross-modal data verification.}
Data matching is a key concept in data integration and data cleaning~\cite{DBLP:books/daglib/0029346}. However, matching and reasoning across different data modalities are not well-addressed.

{\bf (C3)} \textit{Trust of heterogeneous datasets in multi-modal data lakes.}
Evaluating the trustworthiness of web sources for knowledge fusion has been well studied~\cite{DBLP:journals/pvldb/DongGMDHLSZ15}. However, evaluating the trustworthiness of different datasets in data lakes, particularly when they are not well curated, remains an open problem.

{\bf (C4)} \textit{Provenance of the verification process.}
Fully automated verification of generative AI outputs can be challenging. It is important to store the lineage of the end-to-end verification process, in case the retrieved data from data lakes is flawed or incomplete, or the verification process itself makes mistakes. This allows for later human checks or debugging.

\begin{figure}[!t]
    \centering 
    \includegraphics[width=0.9\columnwidth]{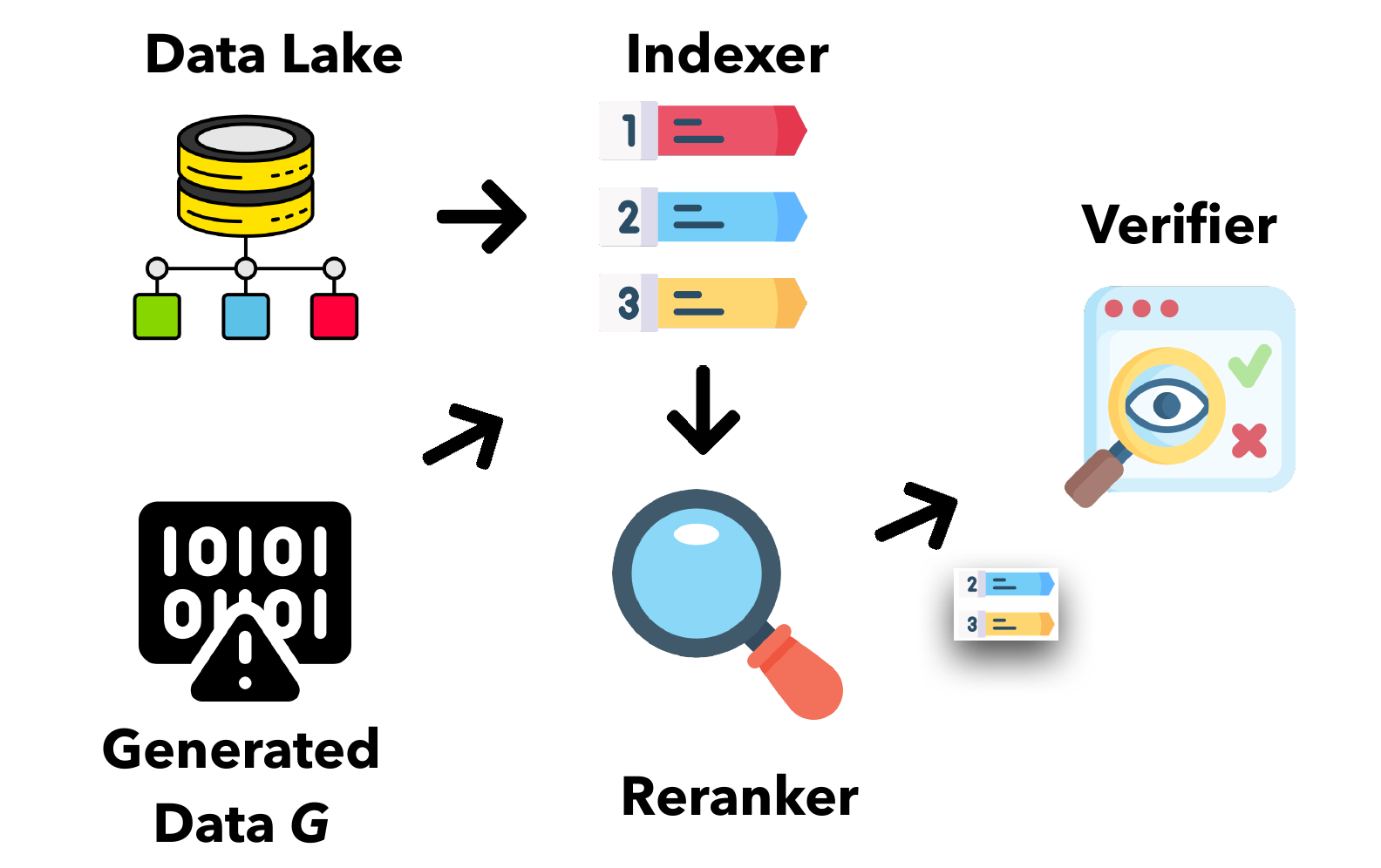}
    \vspace{-1.5em}
    \caption{An overview of \sys.}
    \label{fig:overview} 
    \vspace{-2em}
 \end{figure}

\stitle{Contributions.}
In this paper, we present \sys, a framework for verified generative AI, which offers the following contributions:

\sstab
(1) \textbf{\textit{A modularized framework}}.
We propose a modular framework for verifying generative data that is extendable to different types of data sources and generative data. The framework comprises three main components: an \att{Indexer}, a \att{Reranker}, and a \att{Verifier}, as illustrated in Figure~\ref{fig:overview} (more details can be found in Section~\ref{sec:design}).

The \att{Indexer} module serves the purpose of indexing datasets from diverse modalities including but not limited to tuples, tables, text, and knowledge graphs. It comprises generic and coarse-grained indexes that aid in this task.

The \att{Reranker} module is responsible for reranking the top-$k$ data sources from the \att{Indexer} with respect to a specific generated data object. This step is more fine-grained and task-specific in nature, in order to further optimize the ranking of the retrieved data sources.

For each retrieved data instance, the \att{Verifier} module will determine whether it can verify or invalidate the generated data object.

\sstab
(2) \textbf{\textit{Experiments}}.
We show that \sys achieves high accuracy for generated tables and text (Section~\ref{sec:usecases}), which demonstrates the feasibility of using multi-modal data lakes to verify generative AI.

\sstab 
(3) \textbf{\textit{Open problems}}.
Contribution (1) highlighted the potential of \sys in addressing challenges {\bf C1} and {\bf C2}. However, there are still open problems for these challenges, as well as for other challenges such as {\bf C3} and {\bf C4}, which will be discussed in Section~\ref{sec:conclusion}.
\section{Verified Generative AI: The Problem}
\label{sec:problem}

\stitle{Generative Data.}
Generative data generally refers to data that is created or generated by a model or algorithm, rather than being directly observed or collected from the real world.

In the context of this work, we specifically focus on data generated by large language models, such as \chat, using natural language generation techniques. This includes text data such as paragraphs and sentences, or tabular data such as tuples and tables, but not other data modalities such as image data or graph data.

\stitle{Multi-Modal Data Lakes.}
A multi-modal data lake provides a single repository or data store for storing and managing multiple types of data, including structured, semi-structured, and unstructured data. This can include tables, text, knowledge graphs, images, audio, and other forms of digital content.

This work focuses on showcasing data lakes that are designed to handle and integrate tabular data and text data.

In the rest of this paper, we will use the term \textbf{\textit{data object}} for generated data and \textbf{\textit{data instance}} for a specific unit of data within a data lake, which can take the form of a tuple, table, or text.

We will briefly discuss potential strategies for supporting and integrating other data modalities (\eg knowledge graphs) into a multi-modal data lake in Section~\ref{sec:conclusion}.

\stitle{Verified Generative AI.}
Given a generated data object $\att{g}$, and a multi-modal data lake $\att{L}$, the problem of \textbf{\textit{verified generative AI}} involves discovering a set of data instances $\att{L}^\att{g}$ from $\att{L}$ relevant to $\att{g}$, and verifying each data instance $\att{x} \in \att{L}^\att{g}$ as a mapping from $(\att{g}, \att{x})$ to a ternary value as $\att{verify}(\att{g}, \att{x}) \ra 0 | 1 | 2$, indicating \att{verified} $|$ \att{refuted} $|$ \att{not related}, respectively.

\begin{figure*}[!t]
    \vspace{-1.5em}
    \centering 
    \includegraphics[width=0.9\textwidth]{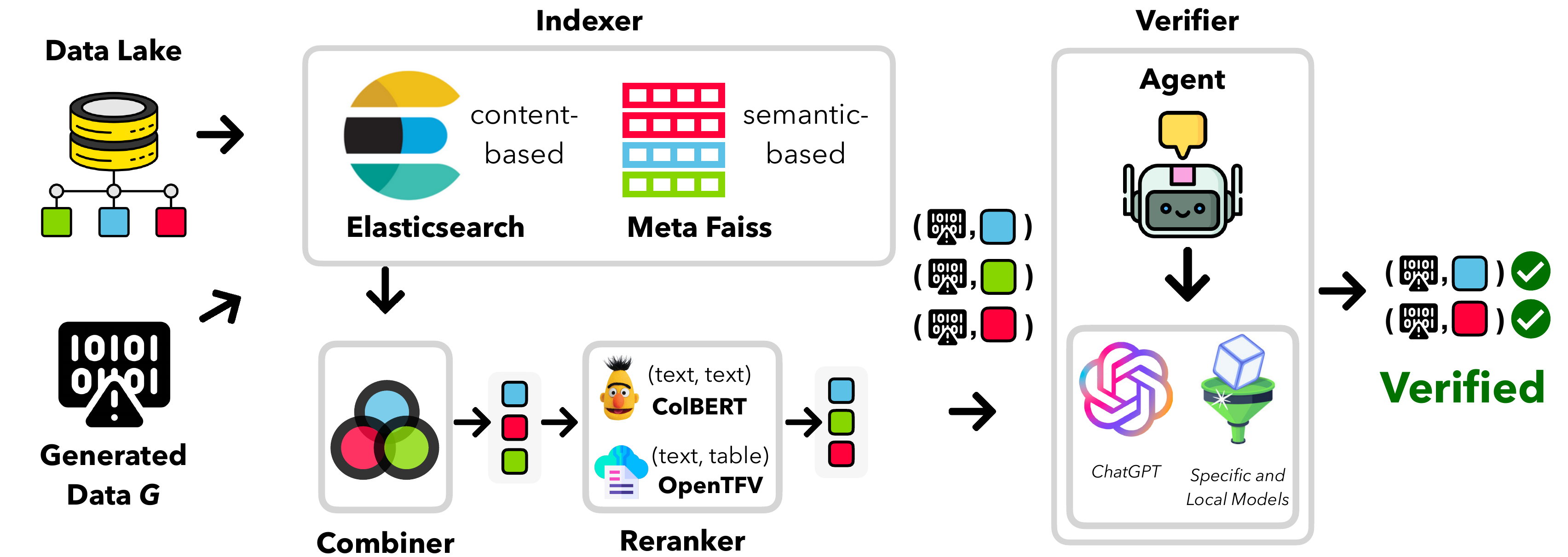}
    \vspace{-1.5em}
    \caption{Modules of \sys.}
    \label{fig:modules} 
    \vspace{-2em}
 \end{figure*}

\etitle{Remark.}
The verification process is specific to the application and requires additional metadata from the user. For instance, when given a tuple, the verification requirement could be either on the entire tuple or on a specific column \eg only verifying the attribute \att{incumbent} in Figure~\ref{fig:examples}(a). 
\section{The Design of VerifAI}
\label{sec:design}

An overview of different modules of \sys is given in Figure~\ref{fig:modules}.

\subsection{Indexer}

The \att{Indexer} module is designed based on two key principles:

\sstab 
(1) {\em Task-agnostic:} 
It can support a wide range of tasks and use cases, making it versatile and adaptable to different needs.

\sstab 
(2) {\em Support for both content- and semantic-based search:} 
It can handle traditional string similarity search as well as vector-based search, enabling more advanced search and retrieval capabilities.

Based on the aforementioned principles, \sys currently consists of two indexes.

\sstab
$\bullet$ Elasticsearch~\cite{elasticsearch} supports content-based search, where tables or text files are serialized as strings and then indexed by Elasticsearch.

\sstab
$\bullet$ Meta Faiss~\cite{faiss} is a library for efficient similarity search and clustering of dense vectors. We first use embedding techniques (\eg tuple-to-vec or text-to-vec using BERT) to convert tuples or chunked text files into vectors, which are then indexed by Faiss.

The above two types of indexes are commonly used for indexing data lakes. For example, Aurum~\cite{DBLP:conf/icde/FernandezAKYMS18} utilizes Elasticsearch, while Microsoft Azure Cognitive Search (\url{https://azure.microsoft.com/en-us/products/search}) indexes embeddings (or vectors) of textual data. Combining these two approaches can enhance recall and serve as a foundation for indexing data lakes more effectively.

\etitle{Remark.}
As these indexes are task-agnostic, the retrieved top-$k$ results may have low accuracy. To ensure the retrieval of correct data instances, $k$ is typically set to a large number (\eg 100 to 1000~\cite{colbert}). However, in order to achieve good accuracy with a much smaller number of results after reranking, it is desired to perform task-aware reranking over the retrieved results.

\stitle{Combiner.}
While different indexes use different techniques (\eg content- or semantic-based,), their retrieved results typically overlap. The \att{Combiner} simply combines these retrieved results from multiple indexes and removes duplicates. 
Due to its simplicity, we did not depict and discuss it in Figure~\ref{fig:overview}.

\subsection{Reranker}

The retrieved results from multiple indexes are initially ranked in a coarse-grained fashion using a task-agnostic similarity measure. However, to achieve a more fine-grained and task-specific ranking of the results, the \att{Reranker} module is utilized.
The primary idea is that, following the reranking process, we only need to focus on a limited number of top-$k'$ retrieved results (\eg $k'=5$).

Next, we describe two different approaches, one for reranking (text, text) pairs, and the other for reranking (text, table) pairs.

$-$ \att{(text, text) Reranker}: We adopt ColBERT~\cite{colbert}, a holistic approach to compare each token of a query and each token of a retrieved text file, resulting in a more precise score.

$-$ \att{(text, table) Reranker}: We have devised a semantic-based reranking method between a natural language statement and a table, called OpenTFV~\cite{DBLP:conf/sigmod/GuFZ0F022}, for open domain table-based fact verification.

\etitle{Remark.}
We are currently working on expanding our support for different types of fine-grained \att{Rerankers}, such as those used for reranking (tuple, tuple) or (text, tuple) pairs.

\subsection{Verifier}

This reasoning module in our system determines whether generated data is verified or refuted, based on a retrieved data instance. It utilizes multiple \att{Verifiers}, each tailored to a specific task. An \att{Agent} decides which \att{Verifier} to use for a given  task.

In \sys, we utilize two types of \att{Verifiers}. The first type is a one-size-fits-all model such as \chat. The second type is composed of specific and localized models designed for different tasks, such as the table fact verification model~\cite{pasta} for (text, table) verification, and a fine-tuned RoBERTa model for (tuple, tuple) verification.

While \chat can be used by default for simplicity, there are two main reasons why we support specific and localized models.

\sstab 
$(1)$ \textit{\textbf{Data privacy.}} 
	Many applications (\eg healthcare, government) contain highly sensitive information. One legitimate concern when using externally hosted models such as \chat is data privacy. The model could potentially learn and retain sensitive information from the data it has seen~\cite{samsung}.

\sstab 
$(2)$ \textit{\textbf{Better accuracy.}}
	Specific and local models, when being fine-tuned on specific datasets and tasks, can oftentimes outperform generic models such as \chat, as will be shown later in Section~\ref{sec:usecases}.

\etitle{Remark.} 
When retrieving data for a generated data instance, it's possible that we may retrieve multiple data instances that either verify or refute the generated data object. This can occur for several reasons, \eg incorrect data instances being retrieved. 
 Therefore, understanding the trustworthiness of different data sources and maintaining the provenance of verification are important.
\section{Preliminary Results}
\label{sec:usecases}

In this section, we showcase preliminary experimental results that highlight the initial achievements of \sys in facilitating the verification of generative AI.

\stitle{Setting.}
We consider three verification tasks.

\begin{center}
{\small
\begin{tabular}{|c|c|}
\hline 
 \rowcolor{black} \textcolor{white}{Generative AI Task} & \textcolor{white}{Retrieved Data Modalities} \\
\hline
 tuple completion & tuples \\
 tuple completion  &  textual files \\
 text generation & relational tables \\
 \hline
\end{tabular}
} 
\end{center}

Note that the (text, text) verification problem is essentially equivalent to the standard fact-checking problem in the natural language processing community [29], which has already been demonstrated to be viable. Therefore, for the sake of this discussion, we will focus primarily on scenarios that involve tuples or tables.

\sstab
$\bullet$ \textbf{\textit{Tuples in need of verification.}}
We collected 100 tuples from web tables. For each tuple, we randomly removed a non-key attribute cell value and then asked \chat to infer the missing value by utilizing the given template provided below. If multiple tuples share the same schema, we can handle them together as a batch.

\begin{center}
{\small
\begin{tabular}{|l|l|l|l|}
  \hline
  \rowcolor{black}\multicolumn{4}{|l|}
  {\textcolor{white}{Prompt template of tuple completion with \chat}} \tabularnewline
  \hline 
  \multicolumn{4}{|l|}
  {\textbf{Question:}} \\
  \multicolumn{4}{|l|}
  {Table name} \\
  column 1 & column 2 & $\ldots$ & column n \\  
  $a_1$ & \at{NaN} & $\ldots$ & $z_1$ \\
  $a_2$ & $b_2$ & $\ldots$ & \at{NaN} \\
  $\ldots$ & $\ldots$ & $\ldots$ & $\ldots$ \\
  \multicolumn{4}{|l|}
  {Please fill the missing values, annotated by \at{NaN}} \\
  \hline
\end{tabular}
} 
\end{center}

\chat will produce a complete table that doesn't have any missing values. An example of this is shown in Figure~\ref{fig:examples}(a). However, each tuple that has had missing values imputed must be verified.

\sstab
$\bullet$ \textbf{\textit{Textual claims in need of verification.}}
We perform a controlled study to assess textual claims, employing 1,300 textual claims from the TabFact~\cite{chen2019tabfact} benchmark, which is currently the most advanced benchmark for verifying the credibility of textual hypotheses by utilizing a given table. This benchmark is suitable for our aim of verifying textual claims. Nevertheless, it's worth mentioning that although TabFact offers a (textual claim, table) pair, our task demands us to search the data lakes for table(s), which is an exceedingly more difficult task.

\stitle{Multi-Modal Data Lakes.}
We utilized 19,498 tables that contain 269,622 tuples, and 13,796 textual files. Of these, 16,573 tables were taken from TabFact~\cite{chen2019tabfact}, while the others were sourced from WikiTable-TURL~\cite{deng2022turl}. Furthermore, since some cells contain references to entities, such as links to Wikipedia pages, we gathered 13,796 entities and retrieved their corresponding text from the associated Wikipedia pages to obtain the textual files.

\stitle{Verifier.}
\chat can serve as the default \att{Verifier} for both data types. The template employed to interact with \chat during the verification process is presented below.

\begin{center}
{\small
\begin{tabular}{|l|l|l|l|}
  \hline
  \rowcolor{black}\multicolumn{4}{|l|}
  {\textcolor{white}{Prompt template of verified with \chat}} \tabularnewline
  \hline 
  \multicolumn{4}{|l|}
  {Please use the evidence below to validate the generative data.} \\ 
  \multicolumn{4}{|l|}
  {\textbf{Evidence:}  \add{[Use the retrieved tuple/table/text]}}  \\
  \multicolumn{4}{|l|}
  {\textbf{Generative Data:}  \add{[Data object to be verified]}}  \\
  \multicolumn{4}{|l|}
  {\textbf{Result:}  Verified/Refuted/Not Related + Further explanation}  \\  
  \hline
\end{tabular}
} 
\end{center}

It is worth mentioning that we have also developed specific and localized models for certain scenarios.
In the case of evaluating (tuple, tuple) pairs, the local model's accuracy is comparable to \chat~\cite{retclean}; therefore, we only present \chat's results.
For (text, table) verification, we employed our previous work PASTA~\cite{pasta} as an alternative to \chat. Later in the section, we will show that PASTA outperforms \chat for (text, table) verification.

\stitle{Evaluation Metrics for Retrieval.} We use Elasticsearch~\cite{elasticsearch} to retrieve the top-3 tuples and top-3 text files for each tuple with an imputed missing value, top-5 tables for each textual claim. As we have a small number of relevant data, we evaluate the retrieval process using only the recall metric.

However, defining ``relevance'' presents a challenge in evaluating the performance of retrieval. Ideally, we would consider all data instances in the data lakes that can support or refute the generative data to be relevant. However, manually labeling relevant data for each tuple or claim would be time-consuming and impractical. Therefore, we take a different approach in this paper.
As each tuple has an original, complete counterpart in the data lake, we consider it to be relevant evidence, while the remaining tuples are considered irrelevant. Moreover, we consider the textual files about entities present in a tuple to be relevant evidence as well.
Additionally, each textual claim is associated with a corresponding table in the original dataset. Thus, we consider the corresponding table to be relevant evidence, while the remaining tables are considered irrelevant.

Due to space limitations, we will not report the effectiveness of reranking in our study. However, it's worth noting that reranking (text, text) pairs has been proven effective in ColBERT~\cite{colbert}, and reranking (tuple, tuple) pairs has been discussed in RetClean~\cite{retclean}.

\stitle{Evaluation Metric for Verifier.} For evaluating the \att{Verifier}'s performance, we use accuracy as the measure.  A \att{Verifier}'s decision is considered correct in one of the following three cases:

\sstab
(1) When the retrieved data instance supports the imputed tuple or claim, the \att{Verifier} outputs ``true'';

\sstab
(2) When the retrieved table refutes the imputed tuple or claim, the \att{Verifier} outputs ``false'';

\sstab
(3) When the retrieved data instance can neither support nor refute it, the \att{Verifier} outputs ``not related''. To compare \chat with PASTA that only offers two different answers: ``true'' or ``false'', in this case, we consider it's also correct when PASTA outputs ``false''.

\stitle{Results.}
The accuracy of \chat in imputing missing values for tuples and determining the correctness of claims is only \textbf{0.52} and \textbf{0.54}, respectively, in the absence of additional data. These findings emphasize the significance of verifying generative data to guarantee accuracy and dependability.

Table~\ref{tab:retrieval_res} displays the retrieval outcomes, revealing that the retrieval module performs well for (tuple, tuple) and (textual claim, table). However, it doesn't retrieve the associated text files effectively based on a tuple. This is due to the fact that we simply utilized Elasticsearch as the \att{Indexer} and only retrieved three text files. We anticipate that the retrieval performance will improve when we expand the number of retrieved files and conduct further experiments by adding the \att{Reranker}.

Table~\ref{tab:verification_res} reveals that as the \att{Verifier}, \chat can accurately determine whether the imputed value is correct or not, with a high accuracy of \textbf{0.88}.
Regarding textual claims, we present results in two settings. When a relevant table is retrieved and provided as evidence to the \att{Verifier} in the form of a (text, relevant table) pair, PASTA is able to verify the textual claim with higher accuracy than \chat based on the table.
However, many retrieved tables are irrelevant to the claim. In all (text, retrieved table) pairs, PASTA's accuracy drops to 0.72 because it hasn't encountered this scenario during training. On the other hand, \chat has superior generalization capabilities and performs better than PASTA when dealing with many irrelevant tables.

\begin{figure}[!t]
  \centering 
  \includegraphics[width=0.95\columnwidth]{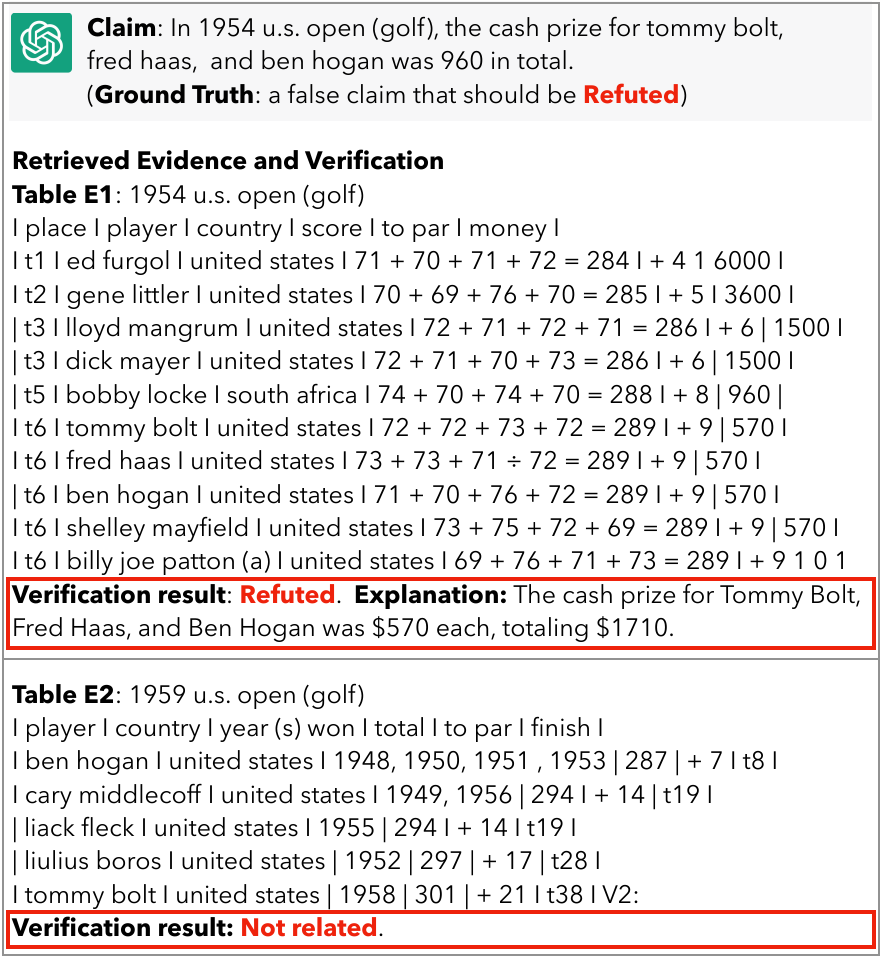}
  \vspace{-1em}
  \caption{Verifying a textual claim using retrieved tables.}
  \label{fig:claim_case} 
  \vspace{-1em}
\end{figure}

In Figure~\ref{fig:claim_case}, we present a case of verifying a textual claim based on retrieved tables using \chat. \sys retrieves two tables $E_1$ and $E_2$, where $E_1$ can be used with an aggregation query to refute the claim while $E_2$ is not related because it is for the year 1959. The red boxes in Figure~\ref{fig:claim_case} show that \chat can provide not only a verification result but also some explanation. Hence, when the retrieved data is highly related to the generative data, local models like PASTA have higher accuracy while protecting privacy. In contrast, \chat is better at generalizing and providing explanations for further judgments. Users can select the appropriate model based on their requirements.

In summary, we have evaluated the usefulness of \sys through various experiments and demonstrated several use cases in Figures~\ref{fig:examples} and \ref{fig:claim_case}.

\begin{table}[t!]
\vspace{-1em}
{\small
\begin{tabular}{|c|c|c|}
\hline
\textbf{Generated data type} & \textbf{retrieved data type} & \textbf{recall}\\ 
\hline \hline
\multirow{2}*{tuple} & tuple & 0.99 \\
\cline{3-3}
& text & 0.58\\ \hline
textual claim & table & 0.88 \\ \hline
\end{tabular}
} 
\caption{Recall on retrieved data instances.}
\label{tab:retrieval_res}
\vspace{-2em}
\end{table}

\begin{table}[t!]
{\small
\begin{tabular}{|c|c|c|c|}
\hline
 & \att{\chat}  & \textbf{PASTA} \\ \hline \hline
(tuple, tuple+text) & 0.88 & NA \\ \hline
(text, relevant table) & 0.75 & \textbf{0.89} \\
(text, retrieved table) & \textbf{0.91} & 0.72\\ \hline
\end{tabular}
} 
\caption{Evaluation on Verifier.}
\label{tab:verification_res}
\vspace{-2em}
\end{table}
\section{Open Problems}
\label{sec:conclusion}

In this paper, we propose a framework called \sys that addresses the growing concern about the reliability of generative AI, which is leading to the spread of misinformation at an alarming rate. The modular design of \sys enables the verification of generated data using multi-modal data lakes, paving the way for research activities that will produce practical solutions for enhancing the reliability of generative AI.
In addition, we have identified important open problems that can make significant advances in this field and improve the reliability of generative AI.

\sstab
$\bullet$ \textbf{\textit{Cross-Modal Data Discovery.}}
Data discovery is a challenging problem in data preparation~\cite{DBLP:conf/icde/FernandezAKYMS18}, particularly in data lakes that contain heterogeneous data stored in various formats, including structured (\eg tables), semi-structured (\eg graphs), and unstructured data. Unlike data lakes that contain only relational tables, discovering data from different modalities requires addressing the heterogeneity of the data. To address this issue, a promising direction is to explore {\em cross-modal representation learning}, which involves encoding data from different modalities into a homogeneous vector space. This approach can facilitate a unified data discovery process, such as using a semantic-based indexer, as illustrated in Figure~\ref{fig:modules}.

\sstab
$\bullet$ \textbf{\textit{Cross-Modal Verification.}}
In addition to textual and relational data, datasets in other modalities, such as knowledge graph entities (or small subgraphs), can contain valuable information for verifying generative AI. As discussed earlier, generic models like \chat may not provide a comprehensive solution for reasoning due to challenges such as privacy and accuracy. Therefore, a promising direction is to develop local models that are specifically trained for certain use cases, such as (text, knowledge graph entity) or (tuple, text). By focusing on specific cases, these local models can provide more accurate and effective solutions for verifying generative AI.

\sstab
$\bullet$ \textbf{\textit{Trustworthiness of Data Sources.}}
The accuracy of discovering and verifying data across different modalities in a data lake can be influenced by the quality and reliability of the underlying data sources. Therefore, it is crucial to assess the trustworthiness of different sources accurately to enhance the overall accuracy and reliability of the entire verification process.

\sstab
$\bullet$ \textbf{\textit{Provenance Management.}}
It is important to maintain a record of the provenance of data instances or sources used in verification to facilitate further human checking or debugging.

\sstab
$\bullet$ \textbf{\textit{Managing Data Generated by Generative AI.}}
Generative AI solution providers, including OpenAI, are continuously collecting prompts and generated data. While this information is valuable for improving generative AI models, it can also be useful for end-users, particularly enterprise users. Therefore, a promising direction is to explore how to manage the (conversational) prompts and data generated by generative AI to enable better prompt engineering and data lineage tracking, similar to ModelDB~\cite{modeldb} which is used for managing machine learning models. 


\balance
\bibliographystyle{ACM-Reference-Format}
\bibliography{citations/ref}

\end{document}